\documentclass[twocolumn,showpacs,preprintnumbers,amsmath,amssymb]{revtex4}
\usepackage{graphicx}% Include figure files
\usepackage{epstopdf}
\usepackage{dcolumn}% Align table columns on decimal point
\usepackage{bm}% bold math

\begin{document}

\title{Neutron Stars Properties and Crust Movements  in Post-glitch  Epoch}
% Force line breaks with \\

\author{L. M. Gonz\'alez-Romero}
\affiliation{Dept. F\'{\i}sica Te\'orica II, Facultad de  Ciencias F\' \i sicas,\\
Universidad Complutense de Madrid, 28040-Madrid, SPAIN}
\author{F. Navarro-L\'erida}
\affiliation{Dept. F\'{\i}sica At\'omica, Molecular y Nuclear, Facultad de  Ciencias F\' \i sicas,\\
Universidad Complutense de Madrid, 28040-Madrid, SPAIN}


\begin{abstract}
Using a new numerical code with non-uniform adapted mesh, we study the changes produced in the global properties of neutron stars by the motion of matter in crust
region  during post-glitch epoch. Our numerical analysis shows that these changes may  contribute to explain the
observed spin-down of rotational  frequency.
\end{abstract}

\maketitle

%\keywords{methods: numerical --- pulsars: general --- stars: neutron --- stars: rotation}

%\section{Introduction}
The sudden spin jumps in rotational frequencies or ``glitches"
observed in many pulsars \citep{exp},
as well as the transient
phase ``post-glitch" returning to the continuous spin-down
process, are the main features of the pulsar evolution and provide
great information sources to study the structure and evolution of
neutron stars \citep{model1}.

The core of  neutron stars is mainly composed of superfluid
neutrons, protons, and  relativistic degenerate electrons (protons
and electrons are found in a small fraction of the neutron
abundance). In the outer kilometer protons are trapped in a
lattice of neutron-rich nuclei (the ``crust"). The region between
nuclei is filled by neutron superfluid \citep{Neutron}.

The crust and the core of the neutron star are strongly coupled
and rotate  with an angular velocity $\Omega_c$. On the other
hand, neutron superfluid, in order to rotate with an angular
velocity $\Omega_s$, establishes an array of quantized vortex
lines parallel to the neutron spin axis (these vortex lines are
surrounded by a standard matter core) \citep{Baym,Alpar1}. 
It is thought that ``glitches" represent a
variable coupling between the core crust component and the neutron
superfluid \citep{Baym}.

The continuous spin-down due to the angular momentum loss, via
electromagnetic radiation caused by the magnetic torque, produces
stresses in the crust that eventually can be broken and brought to
its more spherical equilibrium configuration. The decrease of the
moment of inertia produces a sudden spin jump, which may explain
small glitches \citep{Ruderman1} but not the large (giant) ones
observed in Vela and other pulsars.

Several models use the superfluid as a reservoir of angular
momentum by the mechanism of vortex pinning and unpinning.
Vortices can pin to the nuclei lattice in the inner crust
\citep{Anderson}, not allowing for a radial component in their
velocity. As a consequence, the superfluid stores  angular
momentum \citep{Alpar1}. A sudden unpinning of vortices and their
subsequent motion in the outer direction produce a rapid spin-down
in the superfluid. As the angular momentum can be considered
constant, this also produces a sudden spin-up of the core-crust
component of the star, which gives rise to the observed glitch.

Different models have been proposed in order to explain the glitch
 phenomena: mechanical glitches \citep{Ruderman2}, thermal
glitches \citep{Link},  core driven glitches \citep{Ruderman3},
glitches based on centrifugal buoyancy forces \citep{Carter},
catastrophic unpinning \citep{Cheng}, or annihilation of proton flux
tubes in the crust core boundary \citep{Sedra}. In some of these models
we found a motion of matter  away from  the rotation axis.

In some models the post-glitch transitory epoch is thought to
correspond to a vortex repinning period. In order to explain the
exponential and linear behavior found from fits of  experimental
data of pulsar frequency, Alpar et al (1984, 1993)
have considered the existence of different regions in the crust
with a different behavior in the unpinning process. They
differentiate the regions with  vortex motion in the glitch and
those without vortex motion. Also, the phonon-vortex interaction
has been considered to explain the post-glitch spin-down
\citep{Jones}. Another model  is the crust-cracking model
developed by Franco et al (2000), with the motion of matter
to higher latitudes in the crust induced by starquakes. Recently,
other model for the post-glitch epoch, based on a model for
glitches by Ruderman et al (1998), has been proposed
\citep{Jones2}.

\begin{figure}
\includegraphics[height=6.0cm]{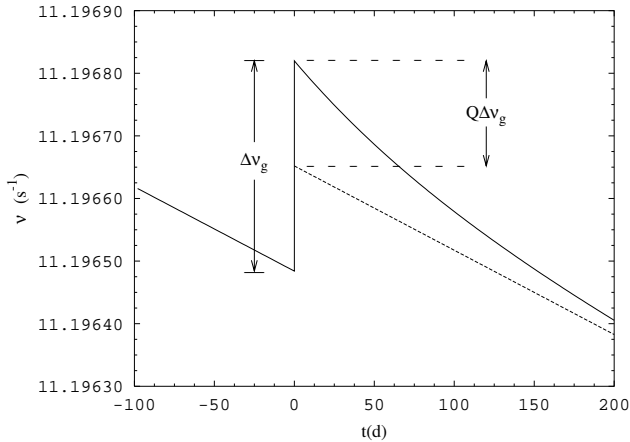}% Here is how to import EPS art
\caption{\label{fig:frequency} Frequency evolution in a typical
large glitch in a pulsar of the Vela type.}
\end{figure}

%\section{The Model}

In this paper, we will study the evolution of a neutron
star in the post-glitch epoch. In order to do that  we will obtain
a family of hydrostatic equilibrium  configurations  by solving numerically
the partial differential equations and the corresponding boundary
conditions for different times in this epoch. We will use a mesh
adapted  to the boundary of the star to avoid the high frequency
oscillations (Gibbs phenomenon) observed near the discontinuity
surfaces (in our case the star boundary) in codes with non-adapted
meshes \citep{Numericos}.

We assume that,
just after the crust cracking, the neutron star reaches a
quasi-equilibrium configuration with a shape  corresponding 
to the angular velocity
at the glitch epoch (the adapting process is almost instantaneous,
as experimental observations indicate). From this moment and
during the post-glitch (which may take from weeks to months, the
time needed to achieve a  situation allowing for
the formation of a new solid crust), the neutron star takes its
quasi-equilibrium shape, in
 almost an  instantaneous manner, following the evolution of the
rotation frequency and  without the constraining stresses of the
crust, which now  is broken. 

Note that in the problem there are two different charateristic times (the rotation period and  the duration of the post-glitch epoch).

Our model of neutron star is composed by two regions; The core and
the crust region. The star boundary (a surface of
constant gravitational plus centrifugal potential) is completely
free (in general, it does not correspond to an ellipsoid) and it
changes with time. We will make the following approximations: the
core can be described by a constant density perfect fluid (given
the results for the density profile obtained in  the studies that
use physical equations of state this seems to be a good first
approximation), the crust region can be described by a
surface density shell  corresponding to a jump in the normal
derivative of the gravitational potential on the star's boundary
(the crust is a thin layer in the outer part of the neutron star),
and  the  angular velocities of the core and
the the crust  are equal. Also, we impose axial and
equatorial symmetry in our Newtonian study  (a
general relativistic version is under development).

We assume that in the post-glitch epoch there is a motion of
matter in the crust region; in our model this
corresponds to changes in the matter density   in the shell found
in the boundary of the star which represents this region. The
stratification expected in the crust region
\citep{Ruderman2,Jones2} supports this hypothesis. This
matter motion may represent a matter motion related with vortex motion in the repinning
process, which is favored near the poles of the star
\citep{Alpar1}, or crust plates motion to the magnetic poles
\citep{Franco}, or magnetized patches motion \citep{Ruderman2}. In
our study we determine a matter motion in the crust
region compatible, inside our framework, with the frequency
evolution observed (Fig. \ref{fig:frequency}).
This motion of matter in the
crust region produces changes in the globlal properties of the neutron star.  We will analyze this effect and its influence on
glitches description.

For each time, in the post-glitch epoch, we will obtain a matter
distribution in the crust region as well as the
corresponding neutron star hydrostatic equilibrium configuration. The time
evolution of the matter distribution in the crust
region and the rest of the properties of the neutron star, are
determined by fitting the rotation frequency and the angular
momentum of the calculated configurations to the values observed
and predicted, respectively, for these quantities during the
post-glitch epoch.

%\section{Numerical Techniques and Results}
Any hydrostatic equilibrium configuration of the family is obtained by solving
the Euler and Poisson equations in the interior of the neutron
star and the Laplace equation in the exterior region. The star
boundary  is determined by the implicit equation $p(r,\theta)=0$,
where $p(r, \theta)$ is the pressure; note that the boundary has
to be found at the same time that the partial differential
equations are solved (i.e., it is a free boundary problem). We
have to impose the following boundary conditions on the
gravitational potential $U$; $U_{in|S} = U_{out|S}$ and
$\partial_n (U_{in}-U_{out})_{|S} = \sigma (\theta)$, where $\sigma (\theta)$
  describes the  surface density profile in the
crust region and $\partial_n$ means normal derivative
to the surface $S$.

To solve this problem we have developed a numerical code based on
the program CADSOL \citep{Cadsol} (a program used for the 
numerical
solution of elliptic partial differential equations by
Newton-Raphson methods, for instance, it has been successfully
used to study Einstein-Yang-Mills fields \citep{EYM} ). Our code
uses an iterative process to calculate the surface of the star and
the solutions of the equations on a boundary adapted non-uniform
mesh. In each iteration the star boundary is moved and the mesh
re-adapted to the new boundary; the Poisson and Laplace equations
are solved with the new boundary conditions and then the Euler
equation determines a new boundary. The process is repeated up to
the moment that all the equations are verified with a given
precision ($10^{-8}$).

%\section{Results}
Now, let us consider the spin frequency time dependence in a typical
glitch, Fig. 1. The post-glitch epoch is fitted using an
exponential dependence on top of a linear dependence
\citep{Baym} (sometimes better fits are obtained with several
exponential terms; here we use a model with just one
exponential term for simplicity). Then the post-glitch frequency
can be described by the following expression
\begin{equation}
  \nu(t) = \nu_0(t) + \Delta \nu_0 [Q e^{-t/\tau}+ 1 - Q],
\end{equation}
 where
$\nu_0(t)= a t + b$.  The meaning of the parameters appears in
Fig. 1. The actual values of the parameters are fixed by the best
fit of the experimental timing data. Here we use a simulation with
the following parameters
\begin{eqnarray}
&  Q=0.5  \; , \; \tau =100 \ {\rm days}  \; , \; \Delta \nu_0 =
3.35894519466 \  10^{-4} , &   \nonumber \\
&   a = -1.34623296 \  10^{-6} \; , \; b= 11.1964839822. &
 \end{eqnarray}
 These parameters  are of the order of magnitude of
those obtained in a real large glitch. Note that the spin jump
value is slightly over a typical one for a large glitch. This is
due to the fact that even when the precision obtained in our
numerical calculations is excellent, we are in the threshold  to
fit the exponential data of a real glitch. Some changes in our
routines are under development to improve the precision and avoid
this limitation.

\begin{figure}
\includegraphics[height=6.0cm]{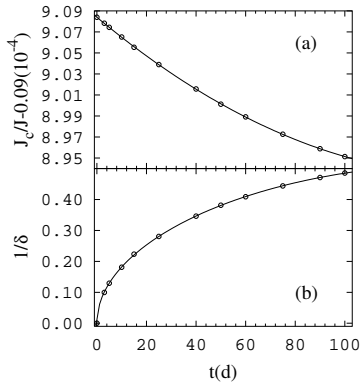}% Here is how to import EPS art
\caption{\label{fig:DJ} Crust-superfluid density parameter and
angular momentum transfer evolution.}
\end{figure}

We concentrate our attention on the  one hundred days just after
the glitch. First, we have to calculate the configuration at the
glitch epoch ($t=0$) after the spin jump. This configuration is
characterized by the spin frequency, core density,
crust-superfluid density (for $t=0$, in a conservative manner, we
use uniform density), and core and crust-superfluid mass. The
shape, angular momentum, and the rest of the physical properties
of the neutron star are then obtained.

The continuous loss of angular momentum, due to the
electromagnetic torque, has also to be considered in the
post-glitch epoch. Here, we assume, in a simplified form, that
this angular momentum loss is proportional to the linear part of
the spin frequency fit (implicitly, we are assuming that in
average the linear changes in the angular velocity of the neutron
star is due, in first approximation, to the electromagnetic
radiation; in general, other effects can contribute to this linear
part \citep{Alpar1}). The constant of proportionality is obtained
from the angular momentum we calculate for the configuration at
the glitch epoch after the spin jump.

Now, in the post-glitch epoch, we have the spin frequency and the
estimated angular momentum of the neutron star for each time $t$.
In order to generate our model we must indicate the form of the
crust-superfluid region surface  mass density.  The effect we want to show
is not very sensitive to the particular choice of the density  profile
 (our model allows a completely general form for the density
profile). We choose a profile for the
crust-superfluid density consisting of a uniform term plus a Gaussian 
distribution centered
 around a latitude near the poles ($\pi/12$ in our case), with relative 
weights $1:0.01$,
\begin{equation}
\sigma = K  \left[1+ \frac{ 0.01 \  \  e^{ -\frac{(\theta
-\pi/12)^2}{2 \delta^2} } }{ \int_0^{\pi/2} e^{-\frac{(t
-\pi/12)^2}{2 \delta^2}} d t } \right].
\end{equation}
Here $K$ and $\delta$ do not depend on $\theta$. (Remark: Concrete values 
$0.01$ and $\pi/12$ have no
special influence on the results described below).
 We also assume constant mass of the neutron star,
constant core density, and constant ratio between the
crust-superfluid mass $M_s$ and the core mass $M_c$,
$\frac{M_s}{M_c} = 0.06$. Then the observed spin frequency
 and the predicted angular momentum of the configuration determine
the shape of the star and the actual crust-superfluid region
density profile.

In order to visualize the physical properties for the
configurations in the family we use two parameters $\delta$ an $o
= \sqrt{1-(R_p/R_e)^2}$ (\textsl{oblaticity}). Note that this
parameter $o$ will not caracterize  completely the shape of the
star, because it is not spheroidal.

\begin{figure}
\includegraphics[height=6.0cm]{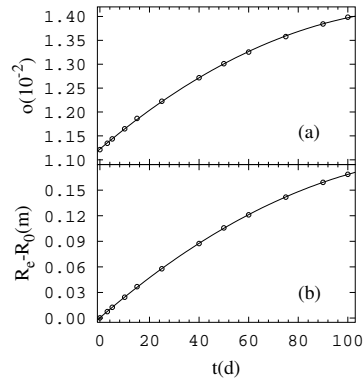}% Here is how to import EPS art
\caption{\label{fig:ReEl} Oblaticity and equatorial radius
evolution.}
\end{figure}

To obtain the values of these two  parameters for each time $t$ we
calculate curves of constant angular momentum in the
two-dimensional parameter space $o-\delta$. In order to do that,
we choose a value for $\delta$ and then we move the oblaticity to
compute an hydrostatic equilibrium configuration with the angular momentum of
the neutron star at the time $t$. The calculated angular velocity
for this configuration will not be in general the actual one of
the neutron star at this time. Owing to that, we have to change
the value of  $\delta$, and repeating the previous algorithm
obtain another configuration. All the process continues by
interpolation method up to the moment the calculated frequency and
angular momentum coincide with those of the neutron star at this
time $t$, within a relative error below $10^{-7}$. We must repeat
this iterative scheme for several values of $t$ during this
post-glitch epoch.

The results for a neutron star with total mass $M=1.4 M_{\odot}$
and initial equatorial radius $R_0 = 12 \ {\rm km}$ are presented
in Figs. 2 and 3. In Fig. 2.b we plot the evolution of the
reciprocal of the density parameter $\delta$. We observe that the
pulsar spin-down, in the post-glitch epoch, represents an increase
of $1/\delta$, which means an increase  of the crust-superfluid
density near the poles (matter approaching the axis). In
principle, this sounds contradictory because a motion of matter to
the axis should decrease the moment of inertia and then a spin-up
is expected. The paradox is solved if we consider the reaction of
the neutron star  to this motion; if we look at the results
presented in Figs 3.a and 3.b, we observe that the star reacts
increasing its oblaticity and equatorial radius, i.e., moving mass
away from the axis. This fact compensates the motion of matter to
the axis in the crust region, which gives rise to the
final result of an increase of the total moment of inertia and the
spin-down of the pulsar. The transfer of angular momentum from the
crust region $J_c$  to the core is presented in Fig. 2.a. The
effect on the frequency of this transfer is overbalanced by the
increased moment of inertia.

%\section{Conclusions}

We have now a new and unexpected effect; the motion of matter to
the axis in the crust region (due to crust plates
motion, vortex repinning or any other origin) produces a reduction
in the pulsar frequency due to the neutron star reaction. In
general, this effect will collaborate with the rest of effects
proposed in the literature to give the final spin-down observed.

Also, we would like to mention that the same effect has been
observed in a constant angular momentum family of configurations,
assuming a motion of matter to the equator in the crust-superfluid
shell. Applying this result to the spin jump in the glitch epoch,
which can be considered as a constant angular momentum phenomenon,
we observe that the neutron star reacts decreasing its oblaticity and
approaching mass to the axis. Then the total moment of inertia is
reduced and the result is a spin up of the neutron star.

Our model have some limitations such as, constant density for the
core and the Newtonian framework. However, we presents for the first time a
global study of neutron star behaviour in the post-glitch epoch
which takes into account the crust region. New numerical 
codes are under development
to improve the model and we are very confident that this new
effect will remain in the scene, collaborating with the rest of
known phenomena to describe glitches, allowing us to have a more
complete picture of pulsar's evolution.

We wish to thank J. Kunz and B. Kleihaus for getting us started on
CADSOL. 

\begin{acknowledgments}

 L.M.G-R. is supported by Spanish Ministry of Education and Science 
Project FIS2005-05198.  F.N-L. is supported by Spanish Ministry of Education and Science 
Project FIS2006-12783-C03-02
\end{acknowledgments}


\begin{thebibliography}{}


\bibitem{exp} V. Radhakrishnan, and R. N. Manchester, Nature, {\bf 222},
228 (1969); P. E. Reichley, and G. S. Downs, Nature {\bf 222}, 229
(1969); P. E. Boynton, E. J. Groth, D. P. Hutchinson, G. P. Nanos,
R. B. Partridge, and D. T. Wilkinson, Astrophys. J. {\bf 175}, 217
(1972); J. M. Cordes, G. S. Downs, and J. Krause-Polstorff,
Astrophys. J. {\bf 330}, 847 (1988); C. S. Flanagan, Nature {\bf
345}, 416 (1990); N. Wang, R. N. Manchester, R. T. Pace, M.
Bailes, V. M. Kaspi, B. W. Stappers, and A. G. Lyne, Mon. Not. R.
Astron. Soc. {\bf 317}, 843 (2000); T. Wong, D. C. Backer, and A.
G. Lyne, Astrophys. J. {\bf 548}, 447 (2001).

\bibitem{model1} C. P. Lorenz, D. G. Ravenhall, and C. J. Pethick,
Phys. Rev. Lett. {\bf 70}, 379 (1993); B. Link, R. I. Epstein, and
J. M. Lattimer, Phys. Rev. Lett. {\bf 83}, 3362, (1999).

\bibitem{Neutron} J. M. Lattimer and M. Prakash, Phys. Rep. {\bf
333}, 121 (2000); H. Heiselberg and V. R. Pandharipande, Ann. Rev.
Nucl. Part. Sci. {\bf 50}, 481 (2000); N. K. Glendenning, {\sl
Compact Stars: nuclear physics, particle physics, and general
relativity}, Astronomy and Astrophysics library, Springer-Verlag,
New York (1996).

\bibitem{Baym} G. Baym, C. J. Pethick, D. Pines, and M. Ruderman,
Nature {\bf 224}, 872 (1969).

\bibitem{Alpar1} M. A. Alpar, P. W. Anderson, D. Pines, and J.
Shaham, Astrophys. J. {\bf 276}, 325 (1984).

\bibitem{Ruderman1} M. Ruderman, Nature {\bf 223}, 597 (1969).

\bibitem{Anderson} P. W. Anderson and N. Itoh, Nature {\bf 256},
25 (1975).

\bibitem{Ruderman2} M. Ruderman, Astrophys. J. {\bf 366}, 261
(1991).

\bibitem{Link} B. Link and R. I. Epstein, Astrophys. J. {\bf 457},
844 (1996); M. B. Larson and B. Link, Mon. Not. R. Astron. Soc.
{\bf 317}, 843 (2000).

\bibitem{Ruderman3} M. Ruderman, T. Zhu, and K. Cheng, Astrophys.
J. {\bf 492}, 267 (1998).

\bibitem{Carter} B. Carter, D. Langlois, and D. M. Sedrakian,
Astron.  Astrophys. {\bf 361}, 795 (2000).

\bibitem{Cheng} K. S. Cheng, D. Pines, M. A. Alpar, and J. Shaham,
Astrophys. J. {\bf 330}, 835 (1988).

\bibitem{Sedra} A. Sedrakian and J. M. Cordes, Mon. Not. R.
Astron. Soc. {\bf 307}, 365 (1999).

\bibitem{Alpar2} M. A. Alpar, H. F. Chau, K. S. Cheng, and D.
Pines, Astrophys. J. {\bf 409}, 345 (1993).

\bibitem{Jones} P. B. Jones, Mon. Not. R. Astron. Soc. {\bf 243},
257 (1990); Astrophys. J. {\bf 373}, 208 (1991).




\bibitem{Franco} L. M. Franco, B. Link, and R. I. Epstein,
Astrophys. J. {\bf 543}, 987 (2000).

\bibitem{Jones2} P. B. Jones, Mon. Not. R. Astron. Soc. {\bf 335},
733 (2002).

\bibitem{Numericos}T. Nozawa, N. Stergioulas, E. Gourgoulhon, and Y. Eriguchi,  Astron. Astrophys., Suppl. Ser.  
 {\bf 132}, 431 (1998).

 \bibitem{Cadsol} W. Sch\"onauer and R. Wei\ss ,
 J. Comput. Appl. Math. {\bf 27}, 279 (1989);
 M. Schauder, R. Wei\ss\ and W. Sch\"onauer,
 The CADSOL Program Package,
 Universit\"at Karlsruhe, Interner Bericht Nr. 46/92 (1992).

\bibitem{EYM} B. Kleihaus, J. Kunz, and  F. Navarro-Lerida, Phys. Rev. Lett.  {\bf 90}, 171101 (2003).













\end{thebibliography}
\end{document}